\documentclass[twocolumn,showpacs]{revtex4}

\usepackage{epsfig}

\newcommand{\beq}{\begin{equation}}
\newcommand{\eeq}{\end{equation}}
\newcommand{\bea}{\begin{eqnarray}}
\newcommand{\eea}{\end{eqnarray}}

\begin{document}
\title{
{\boldmath$B \to \gamma \gamma$} in an ACD model}

\author{I.I.~Bigi $^1$, G.G. Devidze $^2$, A.G. Liparteliani $^2$, and
  U.-G. Mei{\ss}ner $^{3,4}$}

\affiliation{
$^1$  Department of Physics, University of Notre Dame du Lac,
Notre Dame, IN 46556, USA\\ 
$^2$ Institute of High Energy Physics and Informatization, Tbilisi State University, 9 University St., 0186 Tbilisi, Georgia\\
$^3$ Universit\" at Bonn, Helmholtz-Institut f\" ur Strahlen- und Kernphysik (Theorie), D-53115 Bonn, Germany\\ 
$^4$ Institut f\" ur Kernphysik (Theorie), Forschungszentrum J\" ulich, D-52425 J\" ulich, Germany\\
email addresses: ibigi@nd.edu,gela$_{-}$devidze@yahoo.co.uk,lipart48@yahoo.com,
meissner@itkp.uni-bonn.de}

\begin{abstract}
We present a full calculation of the amplitudes for $B_{d[s]} \to \gamma \gamma$ in a simple ACD 
model that extends an incomplete one in a previous paper. We find cancellations between the 
contributions from different KK towers and a small decrease relative to the SM predictions. It is 
conjectured that radiative QCD corrections might actually lead to an enhancement in the branching ratios and {\bf CP} asymmetries, but no more than modest ones. 

\vspace{-6.5cm}

\hfill {\tiny FZJ--IKP(TH)--2008--11, HISKP-TH-08/11, UND-HEP-08-BIG\hspace*{.08em}02}

\vspace{6.5cm}
\end{abstract}

\maketitle

\noindent {\bf 1.}
Exciting times are ahead for fundamental physics, when the LHC experiments start taking data in 2008. 
Many in the community expect a new paradigm to emerge around the TeV scale, be it some variant of SUSY or of Technicolour or  something even more radical, like extra (space) dimensions. Those novel structures can manifest themselves {\em directly} through the production of new quanta or the topology of events or {\em indirectly} by inducing forces that modify rare weak decays. Such indirect searches are not a luxury. We consider it likely that to differentiate between different scenarios of New Physics, one needs to analyze their impact on flavour dynamics. 

In this note we address $B_{d[s]} \to \gamma \gamma$, which has been studied extensively in the Standard Model (SM) and several New Physics scenarios, namely those with non-minimal 
Higgs dynamics and/or SUSY. Within the SM one finds 
\cite{Devidze:2005ua,Lin:1990kw,Simma:1990nr,Herrlich:1991bq}
$$
{\rm BR}(B_s[B_d] \to \gamma \gamma) \simeq 
$$
\beq
10^{-7}\cdot \left(\frac{f_{B_s}}{240\; {\rm MeV}}  \right)^2 
\left[ 10^{-9}\cdot \left(\frac{f_{B_d}}{220\; {\rm MeV}}  \right)^2 \right]  \; . 
\eeq
The exchange of charged scalars in the loop can enhance the 
branching ratios by an order of magnitude \cite{Devidze:1998ub}
. In a previous paper by some of us
\cite{Devidze:2005ua} 
$B_{d,s} \to \gamma \gamma$ has been treated in the ACD model with one extra
dimension \cite{Appelquist:2000nn}
by  calculating the contributions from the Kaluza-Klein (KK) towers of charged scalar fields. 
A small enhancement of a few percent was found there. However there are other contributions as well due to the virtual 
KK towers of Goldstone and of gauge bosons in the loop. Those are computed in this note, which thus 
contains a full evaluation of $B_{d[s]} \to \gamma \gamma$ in a simple ACD model. 

After listing the relevant features of the ACD model in a nutshell we calculate the full ACD contributions to the $B_{d[s]} \to \gamma \gamma$ amplitudes and their {\bf CP} asymmetry. We then give numerical estimates before concluding with an outlook.

\medskip

\noindent {\bf 2.} In the ACD model all particles move in the bulk, i.e. they
are functions of all space-time dimensions. For the bosonic fields one simply
replaces all derivatives and fields of the SM Lagrangian by their
five-dimensional counterparts. The Higgs doublet is chosen to be even
under parity operation in five dimensions and possesses a zero mode. Note
that all zero modes remain massless before the Higgs mechanism becomes
operative and that the fields receive additional masses $\sim n/R$ after
dimensional reduction.  After gauge fixing, one can diagonalize the kinetic
terms for the bosons and derive their propagators. Compared to the SM, there
are additional Kaluza-Klein (KK) mass terms. As they are common to all fields,
their contributions to the gauge boson mass matrix is proportional to the unit
matrix. Because of the KK contribution to the mass matrix, charged and neutral 
Higgs components with $n \neq 0$ no longer play the role of Goldstone bosons.
Instead, they mix with the $W_5^\pm$ and $Z_5$ to form, in addition to the
Goldstone modes $G_{(n)}^0$ and $G_{(n)}^\pm$, three additional physical
states $a_{(n)}^0$ and $a_{(n)}^\pm$. The Lagrangian for the interaction of
the $G_{(n)}^\pm, a_{(n)}^\pm$ and $W_{(n)}$ (the towers of W bosons) with 
ordinary down quarks reads 
\bea
{\mathcal L} &=& \frac{g_2}{\sqrt{2}M_{W(n)}} \bar{Q}_{i(n)} \left(
C_L^{(1)} P_L + C_R^{(1)} P_R \right) \, a^*_{(n)} d_j \nonumber\\
&+& \frac{g_2}{\sqrt{2}M_{W(n)}} \bar{U}_{i(n)} \left(
C_L^{(2)} P_L + C_R^{(2)} P_R \right) \, a^*_{(n)} d_j \nonumber\\
&+& \frac{g_2}{\sqrt{2}M_{W(n)}} \bar{Q}_{i(n)} \left(
C_L^{(3)} P_L + C_R^{(3)} P_R \right) \, G^*_{(n)} d_j \nonumber\\
&+&  \frac{g_2}{\sqrt{2}M_{W(n)}} \bar{U}_{i(n)} \left(
C_L^{(4)} P_L + C_R^{(4)} P_R \right) \, G^*_{(n)} d_j \nonumber\\
&+& \frac{g_2}{\sqrt{2}} \bar{Q}_{i(n)} \, \gamma_\mu
C_L^{(5)} P_L  \, W_{(n)}^\mu d_j \nonumber\\
&+&  \frac{g_2}{\sqrt{2}} \bar{U}_{i(n)} \gamma_\mu
C_L^{(6)} P_L  \, W_{(n)}^\mu d_j
\eea
using the notation of Ref.~\cite{Buras:2002ej}
\bea\label{eq:masspara}
C_L^{(1)} &=& -m_{3}^{(i)} V_{ij} \, , C_L^{(2)} = m_4^{(i)} V_{ij} \, , \nonumber\\
C_R^{(1)} &=& M_{3}^{(i,j)} V_{ij} \, , C_R^{(2)} = -M_4^{(i,j)} V_{ij} \, , \nonumber\\
C_L^{(3)} &=& -m_{1}^{(i)} V_{ij} \, , C_L^{(4)} = m_2^{(i)} V_{ij} \, , \nonumber\\
C_R^{(3)} &=& M_{1}^{(i,j)} V_{ij} \, , C_R^{(4)} = -M_2^{(i,j)} V_{ij} \, , \nonumber\\
C_L^{(5)} &=& c_{(i)n} V_{ij} \, , C_L^{(6)} = -s_{i(n)} V_{ij} \, ,\nonumber\\
M_{W(n)}^2 &=& m^2(a^*_{(n)}) = M^2(G^*_{(n)}) = M_W^2 + \frac{n^2}{R^2}
\eea
with the $V_{ij}$ elements of the CKM matrix. The mass parameters in Eq.~(\ref{eq:masspara})
are defined as
\bea\label{eq:masses}
m_1^{(i)} &=& \frac{n}{R} c_{i(n)} + m_i s_{i(n)} \, , ~
m_2^{(i)}  = -\frac{n}{R} s_{i(n)} + m_i c_{i(n)} \, , \nonumber\\
m_3^{(i)} &=& -M_W  c_{i(n)} + \frac{n}{R} \frac{m_i}{M_W} s_{i(n)} \, , \nonumber\\
m_4^{(i)} &=& M_W  s_{i(n)} + \frac{n}{R} \frac{m_i}{M_W} c_{i(n)} \, , \nonumber\\
M_1^{(i,j)} &=& m_j c_{i(n)} \, , ~~ M_2^{(i,j)} = m_j s_{i(n)} \nonumber\\
M_3^{(i,j)} &=&  \frac{n}{R} \frac{m_j}{M_W} c_{i(n)} \, , 
~~ M_4^{(i,j)} =  \frac{n}{R} \frac{m_j}{M_W} s_{i(n)}~ .
\eea
Here, $M_W$ and the masses of the up (down) quarks $m_i ~(m_j)$ on the
right-hand-side of Eq.~(\ref{eq:masses}) are zero mode masses and the 
$c_{i(n)}, s_{i(n)}$ denote the cosine and sinus of the fermion mixing angles,
respectively: 
\beq
\tan 2\alpha_{f(n)} = \frac{m_f}{n/R}~ , n \geq 1~.
\eeq
The masses for the fermions are given by
\beq
m_{f(n)} = \sqrt{\frac{n^2}{R^2} + m_f^2}~.
\eeq
In what follows, we will utilize the constraint $n/R \geq 250\,$GeV 
\cite{Buras:2003mk} and hence all fermionic angles except $\alpha_{t(n)}$ are 
practically zero.

\medskip

\noindent {\bf 3.} The amplitude for the decay $B_{d[s]} \to \gamma\gamma$ has
the form
\bea\label{eq:amp}
T (B\to \gamma\gamma ) = \varepsilon_1^\mu (k_1)  \varepsilon_2^\mu (k_2) 
\left[ A\, g_{\mu\nu} + i \, B\, \epsilon_{\mu\nu\alpha\beta} k_1^\alpha
  k_2^\beta  \right]\, ,
\eea
where the scalar functions $A \, (B)$ are CP--even(odd). These are calculated
from the 1PR diagrams shown in Fig.~1 with $a_{(n)}, G_{(n)}$ and $W_{(n)}$
particles running in the loops. The corresponding 1PI graphs are suppressed by
$1/M_W^2$ and will be neglected in what follows (see Ref.~\cite{Buras:2003mk}
for more details).
The resulting amplitudes $A_{\rm ACD}$ and $B_{\rm ACD}$
take the from
\bea
A_{\rm ACD} &=& i\frac{\sqrt{2}}{32\pi^2}(eQ_d)^2 f_B G_F
\frac{m_b^3}{m_{s[d]}} \frac{M_W^2}{M^2_{W(n)}} V_{ib}V^*_{is[d]}
\nonumber\\
&\times& \Biggl\{ C(x_{i(n)}) -12\frac{m_i m_{i(n)}}{M_W^2}
c_{(i)n} s_{(i)n} f_1 (x_{i(n)})
\nonumber\\
&-& \frac{3}{2} f_2 (x_{i(n)}) \biggl( 1 + \frac{m_i^2}{M_W^2} -2 
\frac{m_b m_{s[d]}}{M_{W(n)}^2}\frac{n^2}{R^2M_W^2}
\biggr) \Biggr\}
\nonumber\\  
B_{\rm ACD} &=& \frac{2}{m_b^2}  A_{\rm ACD}   
\eea
Where $Q_d$ is charge of down quarks and
\bea\label{fig:C}
C(x) &=& \frac{22x^3-153x^2+159x-46}{6(1-x)^3} +
\frac{3(2-3x)x^2\ln x}{(1-x)^4}\, , \nonumber\\
f_1(x) &=& \frac{5x-3}{6(1-x)^2} + \frac{3x-2}{3(1-x)^3}\ln x ~,
\nonumber\\  
f_2(x) &=& \frac{2x^2+5x-1}{6(1-x)^3} + \frac{x^2}{(1-x)^4}\ln x ~,
\nonumber\\  
x_{i(n)} &=& \frac{m^2_{i(n)}}{M^2_{W(n)}}~.
\eea

\begin{figure}[t]
\vspace*{-1mm}
\centerline{\hspace*{3mm}\epsfig{file=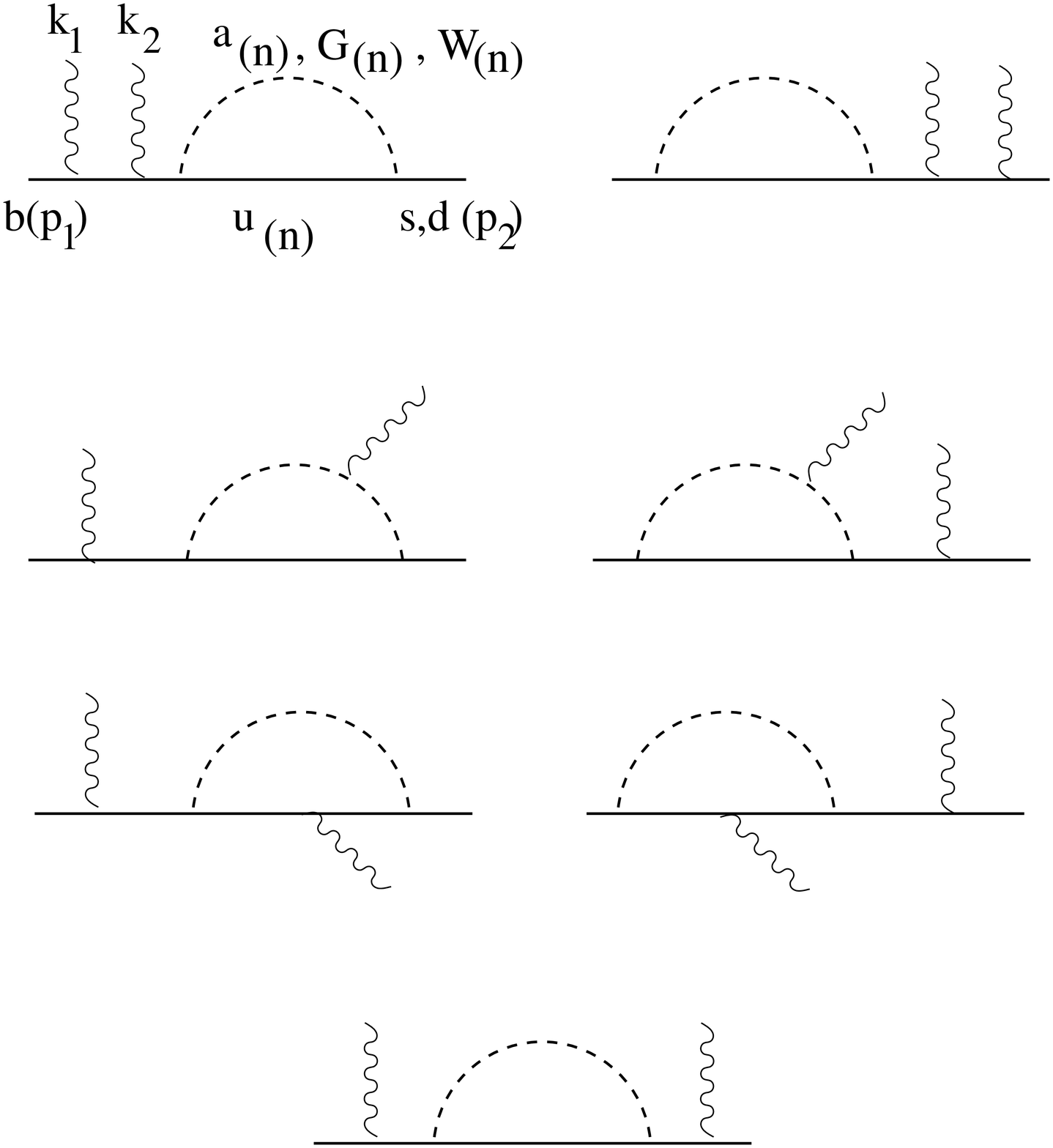,width=5.7cm,height=7.cm}}
\vspace*{-0.5mm}
\caption{1PR diagrams for $B \to \gamma\gamma$ in the ACD model. The dashed lines denote 
the charged KK towers of $a_{(n)}, G_{(n)}$ and $W_{(n)}$, while the solid
line inside the loops represent the up quark KK towers. Wiggly lines 
denote photons and the solid lines in the initial (final) state the
$b$ ($d,s$) quark. Crossed diagrams are not shown. 
}
\label{fig:diag}
\end{figure}

From Eq.~(\ref{eq:amp}) one can readily deduce the expression for the
$B \to \gamma\gamma$ partial decay width $\Gamma$ and the CP asymmetry
parameter $\delta$,
\bea
\Gamma (B \to \gamma\gamma) &=& \frac{1}{32\pi M_B}\left( 4 |A|^2 +
  \frac{1}{2} M_B^4 |B|^2 \right) ~, \nonumber\\
\delta &=& \frac{4|A|^2}{4|A|^2 + M_B^4 |B|^2 /2}~,
\eea
with $M_B$ the corresponding B-meson mass. The asymmetry $\delta$ arises
from the interference of the CP-even and CP-odd parts of the decay amplitude,
see e.g. Refs.~\cite{Herrlich:1991bq,Devidze:1996uw}. The corresponding
expressions for the SM amplitudes $A_{\rm SM}$ and $B_{\rm SM}$ can be taken
from Refs.~\cite{Lin:1990kw,Simma:1990nr}  
\bea
A_{\rm SM} &=& i\frac{\sqrt{2}m_b^3}{32\pi^2 m_{s[d]}} G_F f_B (eQ_d)^2 \lambda_t
\left( C(x_t) + \frac{23}{3}\right)~,\nonumber\\
B_{\rm SM} &=& i\frac{2\sqrt{2}m_b}{32\pi^2 m_{s[d]}} G_F f_B (eQ_d)^2 \lambda_t
\left( C(x_t) + \frac{23}{3} + 16 \frac{m_{s[d]}}{m_b}\right)~,\nonumber\\
&&
\eea
with $G_F$ the Fermi constant, $x_t = m_t^2/M_W^2$ and $C(x)$ as defined in
Eq.~(\ref{fig:C}).

\begin{figure}[t]
\vspace*{-1mm}
\centerline{\hspace*{3mm}\epsfig{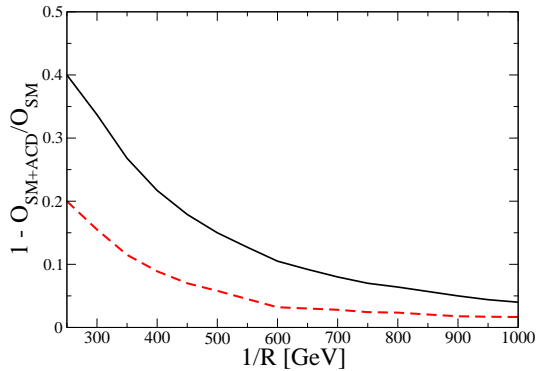}}
\vspace*{-0.5mm}
\caption{B meson partial decay width (solid line) and asymmetry parameter
in the ACD model compared to the SM as a function of the compactification
scale $1/R$.}
\label{fig:ratios}
\end{figure}

We define the relative contribution
of the ACD model with respect to the SM as
\beq
R[{\mathcal O}] = 1 -\frac{{\mathcal O}_{\rm ACD+SM}}{{\mathcal O}_{\rm SM}}~, \quad
({\mathcal O} = \Gamma , \delta )~.
\eeq
The solid line in Fig.~\ref{fig:ratios} shows this ratio for the width and for
the CP-asymmetry parameter. For the smallest compactification radius, $1/R =
250\,$GeV, we have $R[\Gamma] = 0.59$, but with decrasing compactification 
radius the ratio tends towards unity. A similar behaviour is observed for
the asymmetry parameter $\delta$, see the dashed line in  Fig:~\ref{fig:ratios}.

\medskip

\noindent {\bf 4.}
It seems that $B_{d[s]} \to \gamma \gamma$ can realistically be observed only at a Super-B factory that 
for the $B_s$ mode can operate also at the $\Upsilon (5S)$ resonance. The original incomplete evaluation of the ACD contributions showed a rather small enhancement. The full calculation presented 
here revealed a moderate to small 
decrease for the branching ratios and {\bf CP} asymmetries. 
This is not necessarily the last word, though. The QCD corrections yield a substantial enhancement of 20 \% [33 \%] of 
$\Gamma (B_d [B_s] \to \mu^+ \mu^-)$ \cite{Buras:2002ej,Buras:2003mk}. 
Radiative QCD corrections to $B_{d[s]} \to \gamma \gamma$, 
which likewise contains no hadrons in the final state, might be of similar size, yet not the same for the 
$a_{(n)}$, $G_{(n)}$ and $W_{(n)}$ KK towers. Hence we conjecture that the branching ratios of and the 
{\bf CP} asymmetries in $B_{d[s]} \to \gamma \gamma$ can be modified by about 20 - 30 \% in either 
direction, since they can vitiate the cancellations between the different KK contributions we have discussed in this note. Finding an effect of this size in the branching ratio might not be hopeless. 
Maybe the more relevant statement is that the simplest realization of an extra dimension model cannot affect $B_{d[s]} \to \gamma \gamma$ in a numerically large way. If a large deviation from the SM prediction were observed, one had to look elsewhere.

\acknowledgments{
 This work was supported  by the NSF under grant number PHY-0355098 
and by  by Deutsche Forschungsgemeinschaft 
through funds provided to the SFB/TR 16 ``Subnuclear Structure of Matter'',
and by the EU Contract No.~MRTN-CT-2006-035482, ``FLAVIAnet''.
This research is part of the EU Integrated Infrastructure Initiative Hadron
Physics Project under contract number RII3-CT-2004-506078. 
}

\end{document}